\documentclass[aps,pra]{revtex4-1}
\usepackage{amsmath,amssymb}
\usepackage{graphicx}
\usepackage{bm}
\usepackage{bbm}

\newcommand{\rms}{\scriptscriptstyle\rm}

\def\be{\begin{equation}}
\def\ee{\end{equation}}
\def\bea{\begin{eqnarray}}
\def\eea{\end{eqnarray}}

\begin{document}

\title{Bloch-Siegert Shift and its Kramers-Kronig Pair}
\author{Arnab Chakrabarti}
\author{Rangeet Bhattacharyya\thanks{Corresponding author}}
\email{rangeet@iiserkol.ac.in}
\affiliation{Department of Physical Sciences, Indian Institute of Science Education and Research Kolkata, Mohanpur -- 741246, West Bengal, India}

\begin{abstract}
We report that the Bloch-Siegert shift which appears in Nuclear Magnetic
Resonance (NMR) spectroscopy can also be shown to originate as a part of a complex
drive-induced second-order susceptibility term. The shift terms thus obtained are shown to
have an absorptive Kramers-Kronig pair. The theoretical treatment involves a finite
time-propagation of a nuclear spin-$1/2$ system and the spin-bearing molecule under the action of
thermal fluctuations acting on the latter. The finite propagator is constructed to account
for many instances of thermal fluctuations occurring in a time-scale during which the spin
density matrix changes infinitesimally. Following an ensemble average, the resulting
quantum master equation directly yields a finite time-nonlocal complex susceptibility term
from the external drive, which is extremely small but measurable in solution-state NMR
spectroscopy. The dispersive part of this susceptibility term originating from the
non-resonant component of the external drive results in the Bloch-Siegert shift. We have
verified experimentally the existence of the absorptive Kramers-Kronig pair of the
second-order shift term, by using a novel refocussed nutation experiment. Our method
provides a single approach to explain both relaxation phenomena as well as Bloch-Siegert
effect, which have been treated using non-concurrent techniques in the past.
\end{abstract}

\maketitle

\section{Introduction} 

The question that how does a spin system behave in the presence of an external
drive while being connected to a thermal bath, has been investigated in several
notable works, spanning a few decades.  A variety of different approaches
exists in present literature, to deal with different aspects of this problem,
which can be classified into two broad categories. To explain the phenomena of
relaxation and nutation one adopts a quantum master equation approach akin to
Wangsness and Bloch, where only the first order effects of the resonant part of
a weak external drive is considered \cite{wangblo53}. Similar approaches
involve the operation of moving to a doubly-rotating tilted frame before
deriving the master equation as prescribed by Abragam, Vega and Vaughan
\textit{et. al.}, or a heuristic assumption of independent rates of variation
induced by the drive and the relaxation terms with a master equation only for
the latter \cite{abragam06, vega1978, cotandurogryn04}. In all these
approaches, the effect of the non-resonant (counter-rotating) part of an
external drive is ignored. On the contrary, an important feature of the
dynamics of such spin systems is the Bloch-Siegert shift, first reported in a
detailed treatment by Siegert and Bloch, where the counter-rotating (or
non-resonant) terms of the external drive produces a small shift in the
resonance frequency by a factor proportional to $B_1^2$, with $B_1$
defined as amplitude of the drive field \cite{blochsiegert1940}. Such shifts 
follow from some form of perturbation due the external drive while ignoring the
relaxation effects. Later approaches like the Floquet, Magnus or Average
Hamiltonian Theory (AHT) and Fer expansion schemes all employ a perturbative
treatment of the drive while neglecting the relaxation terms \cite{shirley1965,
haeberlen1976, madhu2006}. 

In this work, we strive to develop a single approach whereby both the resonant
and the non-resonant part of a weak drive as well as the relaxation
Hamiltonian, can be treated perturbatively for a spin-$1/2$ system coupled to a
thermal bath undergoing rapid fluctuations. It is expected that the
fluctuations would be present in a heat bath irrespective of the presence of
the coupled spins or in other words, the molecules will undergo collisions
irrespective of whether they bear a spin or not. Hence, we introduce a separate
Hamiltonian which solely acts on the bath as a model of these fluctuations. We
use the method of coarse-graining in time and finite propagation under all
relevant Hamiltonians so as to realize the fact that in the timescales of the
dynamics of the spin system, many instances of the fluctuations take place.  We
find that under these assumptions both the resonant and non-resonant parts of
the drive can be treated perturbatively and the Bloch-Siegert shift naturally
emerges as a second order perturbative correction. More precisely, both the
resonant and non-resonant parts of the drive yield finite, albeit small, second
order correction terms in the form of complex susceptibilities. While the
dominant contribution of the imaginary (dispersive) part of the susceptibility
term takes the form of Bloch-Siegert shift in an asymptotic limit, its
Kramers-Kronig pair yields a decay term. We also note that in the absence of
the coupling to the bath, we recover the expected unitary dynamics of the
spins.

Since the Bloch-Siegert shift is well-studied, we experimentally verify the
existence of the absorptive or the decay term obtained in our method. This
effect is extremely small, and is usually overshadowed by the drive
inhomogeneities. We remove the inhomogeneities by using a novel refocussing
scheme to detect the presence of this additional decay due to the external
drive, in agreement with our theoretical estimates.


\section{Description of the problem} We describe the problem in the context of
solution-state nuclear magnetic resonance spectroscopy, but the arguments can
easily be generalized to other quantum systems coupled to a thermal bath
through a finite set of degrees of freedom. We envisage an ensemble of spin-$1/2$ nuclei
and their respective spin-bearing molecules immersed in a thermal bath (a
liquid solution at a fixed temperature $T$, or inverse temperature $\beta$).
The nuclear spins (henceforth referred to as \emph{system}) interact with the
environment i.e. the bath through the spatial coordinates of the molecules
which cradle the nuclear spins. The molecules are subjected to thermal
collisions with other molecules in the solution.  The entire solution is placed
in a static, homogeneous, magnetic field $\bm{B}_{\circ}$ having magnitude $B_{\circ}$, the direction of which 
is chosen to define our $\hat{z}$ axis.  In the limit of having a thermodynamically large number of spins 
immersed in the bath, the ensemble average of the spin-observables is to be understood as the average
contribution of all spin-bearing molecules present in the bath.

\subsection{Hamiltonians and their timescales} A single spin and its cradle i.e. the molecule (henceforth
referred to as \emph{lattice}) are described by the following Hamiltonian in the laboratory frame,
in the units of angular frequency:
\begin{equation} \label{ham} 
\mathcal{H}(t) = \mathcal{H}^{\circ}_{\rms S} +
\mathcal{H}^{\circ}_{\rms L} + \mathcal{H}_{\rms SL} + \mathcal{H}_{\rms S}(t) + \mathcal{H}_{\rms
L}(t), 
\end{equation} 
where the individual Hamiltonians, their nature and the relevant timescales
are described below.

\begin{itemize} 
\item $\mathcal{H}^{\circ}_{\rms S}$: Zeeman Hamiltonian of the spin-magnetic field
(static) coupling. $\mathcal{H}_{\rms S}^{\circ} = \omega_{\circ}I_z$, 
where $\omega_{\circ} = -\gamma B_{\circ}$ is the Larmor frequency and $\gamma$ is the gyromagnetic ratio of the 
spin-$1/2$ nuclei while $I_{\rms \alpha}$, $\alpha \in \lbrace x,y,z \rbrace$ denote the Cartesian
components of the spin-$1/2$ angular momentum operator. $\omega_{\circ}$ is usually of the order of 100s 
of M-rad/s.  

\item $\mathcal{H}^{\circ}_{\rms L}$: Time-independent Hamiltonian of the cradle (or the
spin-bearing molecule). Since only rotational degrees of freedom are relevant, this Hamiltonian
reflects the molecular rotational energy eigen-system. Rotational (\textit{i.e.} lattice) degrees of freedom
of the ensemble of spin-bearing molecules is assumed to be in equilibrium with the thermal bath at an inverse
temperature $\beta$. This Hamiltonian helps define the thermal
equilibrium  of the molecular rotational part through the equilibrium density matrix
$\rho_{\rms L}^{\rm eq} = \frac{e^{-\beta\mathcal{H}^{\circ}_{\rms L}}}{\mathcal{Z}_{\rms L}}$,
where, $\mathcal{Z}_{\rms L}$ is the partition function.  

\item $\mathcal{H}_{\rms S}(t)$: External transverse drive to the spin system, given by $\mathcal{H}_{\rms S}(t) = 
2\omega_1I_{\rms x}\cos(\omega t)$, where $2\omega_1 = -\gamma B_1$ is the amplitude of the drive Hamiltonian while 
$\omega$ is its frequency. Subsequent analysis will assume a constant $\omega_1$, usually within the range of 1 - 100s 
of kilo-rad$/$s and a constant modulation frequency $\omega$. However, our analysis can easily be generalized to 
$\omega_1(t)$ which is an arbitrary function of time, provided $\omega_1(t)$ is sufficiently slowly varying compared to 
the lattice dynamics. $\omega_1/\omega_{\circ}\sim 10^{-3}$ implying that the spin quantum numbers act as good quantum 
numbers and in principle, $\mathcal{H}_{\rms S}(t)$ can be treated as a perturbation.

\item $\mathcal{H}_{\rms SL}$: Coupling between the spin and the molecular spatial coordinates. It
is usually assumed to be time-independent in the laboratory frame and involves operators of both
spin and molecular (spatial) coordinates. We use a generic form of the coupling Hamiltonian as described by 
Wangsness and Bloch \cite{wangblo53}. Its amplitude is denoted by $\omega_{\rms SL}$. 
We assume a weak coupling to the lattice i.e. $\omega_{\rms SL}/\omega_{\circ}\ll 1$.  

\item $\mathcal{H}_{\rms L}(t)$: Fluctuations in the lattice. It embodies the effect of thermal
collisions experienced by the spin-bearing molecule. Time-scales associated with this process are of
the order of the time scale of molecular collisions which is less or equal to the rotational
correlation time in liquids (of the order of a few picoseconds \cite{samidor85, tanabe78,gilgri72}).
The exact form of these processes can be quite complicated in the time scales of molecular
collisions, but we adopt a simplified model based on the properties of the thermal equilibrium as
described in the following paragraphs.  

\end{itemize}

We model the term due to molecular collisions, $\mathcal{H}_{\rms L}(t)$ as stochastic fluctuations
of the rotational energy levels (energy levels of $\mathcal{H}^{\circ}_{\rms L}$).  Since the
fluctuations do not drive the lattice away from equilibrium, we choose $\mathcal{H}_{\rms L}(t)$ to be
diagonal in the eigen-basis $\lbrace\vert\phi _j\rangle\rbrace$ of $\mathcal{H}^{\circ}_{\rms L}$,
represented by $\mathcal{H}_{\rms L}(t) = \sum_{j}f_j(t)\vert\phi_j\rangle\langle\phi_j\vert$.
$f_j(t)$-s are assumed to be independent, Gaussian, $\delta$-correlated stochastic variables with
zero mean and standard deviation $\kappa$.  

We note that the fluctuations $\mathcal{H}_{\rms L}(t)$, which solely act on the lattice, may assume
different values for different ensemble members at different time instants, while respecting the
constraints imposed by the requirement of sustained thermal equilibrium. On the contrary, the other
terms in the Hamiltonian (\ref{ham}) are identical for all ensemble members.  

Starting with this description, we perform all subsequent calculations in the interaction
representation of $\mathcal{H}^{\circ}_{\rms S} + \mathcal{H}^{\circ}_{\rms L}$ and all Hamiltonians
in this representation are denoted by $H$ with relevant subscripts.  Since $\mathcal{H}_{\rms
L}(t)$ commutes with $\mathcal{H}^{\circ}_{\rms L}$ at all times, the form of the lattice
fluctuations remain unchanged in the interaction representation. 

In the interaction representation, let $\rho_{\rms S}(t)$ denote the density matrix of the spin-$1/2$ ensemble 
and $F(t)$ denote any arbitrary spin-observable for this ensemble. The dynamics of the expectation value of 
$F(t)$, denoted by $M(t) = \text{Tr}_{\rms S}\big[F(t)\rho_{\rms S}(t)\big]$
is given by
\begin{eqnarray}\label{de1}
   \frac{d}{dt}M(t) = \text{Tr}_{\rms S}\Big[\Big\lbrace\frac{d}{dt}F(t)\Big\rbrace\rho_{\rms S}(t)\big] +
                                              \text{Tr}_{\rms S}\Big[F(t)\Big\lbrace\frac{d}{dt}\rho_{\rms S}(t)\Big\rbrace\Big],
\end{eqnarray}
where, $\text{Tr}_{\rms S}$ denotes the trace over the spin-degrees of freedom. Since the spins are coupled to their respective molecules,
$\rho_{\rms S}(t)$ is obtained from the full density matrix of the spin-molecule ensemble, $\rho(t)$ by tracing over the lattice degrees 
of freedom (denoted by $\text{Tr}_{\rms L}$). Hence in order to evaluate the r.h.s.\;of equation (\ref{de1}) we need a suitable expression 
for $\frac{d}{dt}\rho_{\rms S}(t)$, which is obtained from a quantum master equation, derived for the Hamiltonian $\mathcal{H}(t)$. In the
following sections we derive such a master equation and explicitly show the dynamical equations relevant in the context of solution state 
NMR spectroscopy of an ensemble of spin-$1/2$ nuclei.


\section{Master equation with finite propagation for fluctuations}

We seek to derive a master equation which captures the dynamics of the spin system described in the
previous section. Since our problem concerns a single Hilbert space, a part of which undergoes rapid
fluctuations we follow the standard practice of, (i) propagating for a large enough time $\Delta t\;
(> 0)$ over which fluctuations can be adequately averaged out, yet in the same interval $H_{\rms
S}$ and $H_{\rms SL}$ should remain linearizable as is uasually done, (ii) taking ensemble average and a trace
over the lattice variables and (iii) finally using the coarse-grained equation thus obtained as our
dynamical equation with coarse-grained time derivatives replaced by ordinary ones
\cite{cotandurogryn04}. The step (i) requires that the system and the fluctuations have widely
separated timescales of evolution \textit{i.e.} $\tau_c \ll \omega_{\rms 1}^{-1} , \omega_{\rms
SL}^{-1}$,
where $\tau_{\rms c}$ is the time during which the lattice correlations are significant, such that
we can find a $\Delta t$ which obeys 
$\tau_{\rms c} \ll \Delta t \ll \omega_{\rms 1}^{-1}, \omega_{\rms SL}^{-1}$.

We begin from the von-Neumann Liouville equation for a single spin and its cradle, the molecule, 
whose density matrix is denoted by $\widetilde{\rho}(t)$,
\begin{eqnarray}
\frac{d}{dt} \widetilde{\rho}(t) = -i \left[ H(t),\; \widetilde{\rho}(t) \right],
\end{eqnarray}
where $H(t) = H_{\rms S}(t) + H_{\rms SL}(t) + H_{\rms L}(t)$.
The formal solution of the above equation for a finite time interval $t$ to $t+\Delta t$ is given by,
\begin{eqnarray}
\widetilde{\rho}(t+\Delta t) = \widetilde{\rho}(t) -i\int\limits_{t}^{t+\Delta t}\kern-6pt dt_1 \;
\left[ H(t_1),\; \widetilde{\rho}(t_1) \right]
\end{eqnarray}
For the dynamics of the spin (or system) part, we obtain from the above, 
by taking trace over lattice variables,
\begin{eqnarray}
\widetilde{\rho}_{\rms S}(t+\Delta t) &=& {\rm Tr}_{\rms L}\!\left\{\widetilde{\rho}(t+\Delta t)\right\} \nonumber  \\
&=& {\rm Tr}_{\rms L}\!\left\{\widetilde{\rho}(t)\right\} -i\int\limits_{t}^{t+\Delta t}\kern-6pt 
dt_1\; {\rm Tr}_{\rms L}\! 
\left[H_{\rms eff}(t_1) + H_{\rms L}(t_1),\; U(t_1)\widetilde{\rho}(t)U^{\dagger}(t_1)\right]
\nonumber \\
&=& \widetilde{\rho}_{\rms S}(t) -i\int\limits_{t}^{t+\Delta t}\kern-6pt dt_1\; {\rm Tr}_{\rms L}\!
\left[H_{\rms eff}(t_1),\; U(t_1)\widetilde{\rho}(t)U^{\dagger}(t_1)\right], \label{exact}
\end{eqnarray}
where, $H_{\rms eff}(t) = H_{\rms S}(t) + H_{\rms SL}(t)$,
$U(t_1) = U(t_1,t) = T \exp[-i\int_{t}^{t_1}dt_2 H(t_2)]$, and $T$ is the Dyson time-ordering
operator.
In the above, the commutator involving $H_{\rms L}(t_1)$ vanishes due to the partial trace and
$\widetilde{\rho}_{\rms S}(t)$ denotes the single spin density matrix.

To obtain a master equation for the spin system we perform a finite-time propagation of the r.h.s.
of (\ref{exact}) by keeping terms only upto the leading second order in $H_{\rms eff}$ while
retaining all orders of $H_{\rms L}$. We emphasize that this construction is at the immediate next
level of approximations as that of Bloch and Wangsness (barring the fluctuations), where only the
leading linear order of $H_{\rms S}$ was retained while having quadratic orders in $H_{\rms SL}$
\cite{wangblo53}.  Since we intend to capture the dynamics of the spin part while the lattice
undergoes a large number of fluctuation instances, a form of the propagator $U$ is required which
captures the finite propagation due to $H_{\rms L}$ while only retaining the leading order linear
term in $H_{\rms eff}$, in order to capture the overall second order effects due to $H_{\rms eff}$.
Such a form of the propagator is readily available from Neumann series as (Appendix: section A),
\begin{eqnarray}\label{finprop}
U(t_1) \approx U_{\rms L}(t_1) - i\int\limits_{t}^{t_1}\kern-0pt dt_2\; H_{\rms eff}(t_2)\, U_{\rms
L}(t_2)
\end{eqnarray}
with $U_{\rms L}(t_1) = U_{\rms L}(t_1,t) = T \exp[-i\int_{t}^{t_1}dt_2 H_{\rms L}(t_2)]$.
We note that the above truncated form of $U$ is strictly applicable only, (i) if at least a part of $H_{\rms eff}(t)$ does not
commute with $H_{\rms L}(t)$ (\textit{i.e.} $\omega_{\rms SL}\not=0$) and (ii) $H_{\rms eff}(t)$ has a
timescale much slower than the timescale of the fluctuations.
In the case where $\omega_{\rms SL}=0$, we have,
$U(t_1) = U_{\rms S}(t_1)U_{\rms L}(t_1)$ with $U_{\rms S}(t_1) = \exp\left[- i \int_t^{t_1}\kern-0pt dt_2\;
H_{\rms S}(t_2)\right]$,
which results in pure unitary evolution of the spin systems under the external drive in the form of an
infinite Dyson series.  Therefore, setting the coupling between the spin and the lattice to zero,
results in a spin dynamics which is completely decoupled from the lattice dynamics (molecular
collisions and hence the fluctuations). Our use of equation (\ref{finprop}), requires that $\omega_{\rms SL}\not=0$ in the
subsequent calculations. 

We substitute equation (\ref{finprop}) in the equation (\ref{exact}) to obtain,
\begin{eqnarray}
\widetilde{\rho}_{\rms S}(t+\Delta t) &=& \widetilde{\rho}_{\rms S}(t)
-i\int\limits_{t}^{t+\Delta t}\kern-6pt dt_1\; {\rm Tr}_{\rms L}\!\left[H_{\rms eff}(t_1),\;
U_{\rms L}(t_1)\widetilde{\rho}(t)U_{\rms L}^{\dagger}(t_1)\right] \nonumber \\
&-& \int\limits_{t}^{t+\Delta t}\kern-6pt dt_1\; \int\limits_{t}^{t_1}\kern-2pt dt_2\;
{\rm Tr}_{\rms L}\!\left[H_{\rms eff}(t_1),\;
H_{\rms eff}(t_2)\,U_{\rms L}(t_2)\widetilde{\rho}(t)U_{\rms L}^{\dagger}(t_1) 
- U_{\rms L}(t_1)\widetilde{\rho}(t)U_{\rms L}^{\dagger}(t_2)\,H_{\rms eff}(t_2)\right] + 
\mathcal{O}[H_{\rms eff}^{3}]\label{semifinal}
\end{eqnarray}
We note that the above form is exact upto the leading second order in $H_{\rms eff}$ and yet captures evolution
solely under $H_{\rms L}$ upto, in principle, infinite orders through the $U_{\rms L}$ terms.

Next, we perform an ensemble averaging of both sides of the equation (\ref{semifinal}) and neglect 
the third and the higher order contributions of $H_{\rms eff}$.
Assuming that at the beginning of the coarse-graining interval, the density matrix for the whole
ensemble can be factorized into that of the system and the lattice with the latter at thermal
equilibrium, we obtain 
\begin{equation}\label{ddm1} 
\overline{U_{\rms L}(t_1)\widetilde\rho(t)U_{\rms L}^{\dagger}(t_2)} =
\rho_{\rms S}(t)\otimes\rho_{\rms L}^{\rms eq}\exp\big(-\frac{1}{2}\kappa^2\vert t_1 - t_2\vert\big), 
\end{equation} 
where, $\rho_{\rms S}(t)$ denotes the density matrix of the system whereas $\rho_{\rms L}^{\rms eq}$ 
denotes the equilibrium density matrix of the lattice (Appendix: section B). 
Using the
above result we find that the integrands in the second order terms of the coarse-grained equation
(\ref{semifinal}) takes the form of a double commutator decaying within the timescale of
$2/\kappa^2$.  Thus $2/\kappa^2$ forms the upper bound of the timescales during which the lattice
correlations are significant and as such we replace it by $\tau_c$. We thus have an equation of the
form

\begin{eqnarray}\label{m2} 
\rho_{\rms S}(t+\Delta t) - \rho_{\rms S}(t) & = & -i\int\limits_t^{t+\Delta t}\kern-6pt dt_1\;
{\rm Tr}_{\rms L}\!\left[H_{\rms
eff}(t_1),\, \rho_{\rms S}(t)\otimes\rho_{\rms L}^{\rms eq}\right] \nonumber\\ 
&-&\int\limits_t^{t+ \Delta t}\kern-6pt dt_1\;\int\limits_t^{t_1}\kern-2pt dt_2\; 
{\rm Tr}_{\rms L}\!\left[H_{\rm eff}(t_1),\,\left[H_{\rms eff}(t_2), 
\rho_{\rms S}(t)\otimes\rho_{\rms L}^{\rms eq}\right]\right]e^{-{\vert t_1-t_2\vert}/{\tau_c}}.
\end{eqnarray}
Next, following the prescription of Cohen-Tannoudji \textit{et.al.}, we
divide both sides of the resulting equation by $\Delta t$ and approximate the coarse-grained
derivative thus obtained on the l.h.s by an ordinary time derivative \cite{cotandurogryn04}.
Subsequently, we take the limit $\Delta t/\tau_c \rightarrow \infty$ to arrive at
the following master equation,
\begin{eqnarray}\label{mf} 
\frac{d}{dt}\rho_{\rms S}(t) & = & -i\; {\rm Tr}_{\rms L}\!\left[H_{\rms
eff}(t),\, \rho_{\rms S}(t)\otimes\rho_{\rms L}^{\rms eq}\right]^{\rms sec}
- \int\limits_{0}^{\infty}\kern-2pt d\tau\;
{\rm Tr}_{\rms L}\!\left[H_{\rms eff}(t),\,\left[H_{\rms eff}(t-\tau), 
\rho_{\rms S}(t)\otimes\rho_{\rms L}^{\rms eq}\right]\right]^{\rms
sec}e^{-{\vert\tau\vert}/{\tau_c}},
\end{eqnarray} 
where, the superscript `sec' denotes that only the secular contributions are retained (ensured
by the coarse graining) \cite{cotandurogryn04}. Unlike the usual forms of the master equation found in literature, equation (\ref{mf}) has a
finite, time-nonlocal, second order contribution of the external drive to the system \cite{brepet02,
cotandurogryn04, wangblo53, red57}.  The equation (\ref{mf}) yields Lorenztian spectral density functions due to 
the presence of the exponential decay term
and correctly predicts the relaxation behavior
along with the first-order nutation 
of the irradiated spin system as in other forms the quantum master equations \cite{wangblo53, red57, brepet02, cotandurogryn04}. 


\section{Bloch Equations with second-order drive susceptibilities}

With the quantum master equation (\ref{mf}) we can in-principle determine the dynamical equations
for the expectation values of any spin observable using equation (\ref{de1}). In the context
of nuclear magnetic resonance (NMR) spectroscopy we assume that the external drive is nearly resonant
i.e. $\omega = \omega_{\circ} + \Delta\omega$ where $\Delta\omega/\omega_{\circ}\rightarrow 0$. This 
implies that the heterodyne detection followed by low-pass filtering 
in a typical NMR measurement is equivalent to measurements made in a co-rotating frame of frequency $\omega$ \cite{callaghan11}. 
As such the interaction representations of the relevant co-rotating spin-$1/2$ observables are given by,

\begin{align}\label{obsint}
   F_{\rms x}^{\rms R}(t)  & = e^{-i\Delta\omega t I_z}I_xe^{i\Delta\omega t I_z} = \frac{1}{2}\big[ I_+e^{-i\Delta\omega t} + I_-e^{i\Delta\omega t}\big]\nonumber\\
   F_{\rms y}^{\rms R}(t)  & = e^{-i\Delta\omega t I_z}I_ye^{i\Delta\omega t I_z} = \frac{1}{2i}\big[ I_+e^{-i\Delta\omega t} - I_-e^{i\Delta\omega t}\big]\nonumber\\
   F_{\rms z}^{\rms R}(t)  & = e^{-i\Delta\omega t I_z}I_ze^{i\Delta\omega t I_z} =  I_z,
\end{align}
where $I_{\pm} = I_x \pm i I_y$, with the understanding that the expectation values 
of $F_{\rms \alpha}^{\rms R}(t)$, $\alpha \in \lbrace x,y,z \rbrace$ defines the measured $\alpha$-magnetizations, $M_{\rms \alpha}(t)$.

Also the external drive, in the interaction representation, is $H_{\rms S}(t) = \omega_1\big[F_{\rms x}^{\rms C}(t) + F_{\rms x}^{\rms R}(t)\big]$,
where the counter-rotating component of the drive-field is $F_{\rms x}^{\rms C}(t) = \frac{1}{2}\big[ I_+e^{i\Omega t} + I_-e^{-i\Omega t}\big]$ 
with $\Omega = \omega + \omega_{\circ}$. The dynamical equations for the measured magnetization components, $M_{\alpha}(t)$ can now 
be obtained directly from equations (\ref{de1}) and (\ref{mf}) using the observables defined in equation (\ref{obsint}). The near resonance condition 
demands that in the secular limit, only the terms in $H_{\rms S}(t)$ with frequency $\Delta\omega$ (resonant or co-rotating 
terms) i.e. terms in $F_{\rms x}^{\rms R}(t)$, contribute in the first order of equation (\ref{mf}). Following the usual practice, for an isotropic heat bath, 
we assume that $\text{Tr}_{\rms L}[H_{\rms SL}(t),\rho_{\rms S}(t)\otimes\rho_{\rms L}^{\rms eq}] = 0$, which in turn ensures that the cross-terms between 
the drive and the coupling, in equation (\ref{mf}), vanish identically in the second-order \cite{wangblo53, brepet02, cotandurogryn04}. Thus $H_{\rms SL}(t)$ has 
no first order contribution in equation (\ref{mf}) and its second-order contributions lead to the relaxation times $T_1$ and $T_2$ (longitudinal and transverse relaxation
times respectively) as well as the equilibrium magnetization $M_{\circ}$, exactly in the same way as in Wangsness and Bloch's work \cite{wangblo53}. 

On the other hand the second order secular drive terms have contributions from both the resonant ($F_{\rms x}^{\rms R}(t)$) as well as the 
non-resonant ($F_{\rms x}^{\rms C}(t)$) parts, resulting in complex susceptibilities proportional to $\omega_1^2$. The secular integration
in equation (\ref{m2}) i.e. integration over $t_1$, makes the cross-terms between $F_{\rms x}^{\rms R}(t)$ and $F_{\rms x}^{\rms C}(t)$ as 
well as the non-secular self-terms of $F_{\rms x}^{\rms C}(t)$, negligibly small in the second-order \cite{cotandurogryn04}. Thus the master 
equation (\ref{mf}) retains only the secular self-terms of $F_{\rms x}^{\rms C}(t)$ in the second-order of drive-perturbation while retaining all 
possible self-terms from $F_{\rms x}^{\rms R}(t)$, which is manifestly secular.
The absorptive and dispersive components of the second-order drive susceptibilities thus obtained involve Lorentzian spectral-density functions centered at 
$\Delta\omega$ and $\Omega$  and result in additional damping and shift terms in the dynamical equations. 

Explicit calculations for the second-order drive contributions including all the relevant commutation relations can be found in the Appendix: 
section C. Neglecting Lamb-Shift contributions from $H_{\rms SL}(t)$, we then arrive at the following form of the Bloch-equations:

\begin{eqnarray}\label{BE}
     \frac{d}{dt}M_{\rms z}(t) & = & \omega_1M_{\rms y}(t) -\frac{1}{T_1}\big[M_{\rms z}(t) - M_{\circ}\big] -\eta_z\,M_{\rms z}(t)\nonumber\\
     \nonumber\\
     \frac{d}{dt}M_{\rms x}(t) & = & \big[\Delta\omega - \omega_{\rms BS}\big]M_{\rms y}(t) - \frac{1}{T_2}M_{\rms x}(t) - \eta_x\,M_{\rms x}(t)\nonumber\\
     \nonumber\\
     \frac{d}{dt}M_{\rms y}(t) & = & -\big[\Delta\omega - \omega_{\rms BS} -\delta\omega\big]M_{\rms x}(t) -\omega_1M_{\rms z}(t) - \frac{1}{T_2}M_{\rms y}(t) 
                                        - \eta_y\,M_{\rms y}(t),
\end{eqnarray}
where 
\begin{eqnarray}\label{BS1}
     \omega_{\rms BS} = \frac{1}{2}\Big(\frac{\omega_1^2\Omega\tau_c^2}{1 + \Omega^2\tau_c^2}\Big)
\end{eqnarray}
is the frequency-shift originating from $F_{\rms x}^{\rms C}(t)$, while
\begin{eqnarray}\label{shift}
     \delta\omega = \frac{\omega_1^2\Delta\omega\tau_c^2}{1 + \Delta\omega^2\tau_c^2}
\end{eqnarray}
is the same from $F_{\rms x}^{\rms R}(t)$. The damping coefficients are

\begin{eqnarray}\label{damp}
     \eta_z & = & \omega_1^2\Big[\frac{\tau_c}{1 + \Omega^2\tau_c^2} + \frac{\tau_c}{1 + \Delta\omega^2\tau_c^2}\Big]\nonumber\\
     \eta_x & = & \omega_1^2\frac{1}{2}\Big[\frac{\tau_c}{1 + \Omega^2\tau_c^2}\Big] \nonumber\\
     \eta_y & = & \omega_1^2\Big[\frac{1}{2}\Big(\frac{\tau_c}{1 + \Omega^2\tau_c^2}\Big) + \frac{\tau_c}{1 + \Delta\omega^2\tau_c^2}\Big].
\end{eqnarray}

The shift and the damping coefficients thus arrived at, are Kramers-Kronig pairs obtained from the second order drive
susceptibilities. 

We note that in the limit $\Omega \tau_c > 1$, a condition often met in solid state NMR
spectroscopy because of slow fluctuations, $\omega_{\rms BS}$ converges to $\omega_1^2/2\Omega$ 
which we readily identify with the familiar Bloch-Siegert shift. For a more explicit comparison with Bloch and Siegert's original
expression, the condition $\Delta\omega = 0$ introduces a shift in the resonance field given by

\begin{eqnarray}\label{BS2}
      -\frac{1}{\gamma}\omega_{\rms BS} = -\frac{1}{\gamma}\frac{\omega_1^2}{4\omega_{\circ}} = \frac{B_1^2}{16 B_{\circ}},
\end{eqnarray}
in this limit \cite{blochsiegert1940}. Also, $\Delta\omega = 0$ implies $\delta\omega = 0$ and the only frequency shift term arises
from the counter-rotating component of the external drive i.e. $\omega_{\rms BS}$. Since $\Delta\omega$ is small
in magnetic resonance experiments and $\tau_c\ll \omega_1^{-1}$, $\delta\omega$ is negligible for all practical purposes.

On the other hand when $\Omega\tau_c < 1$, the absorptive terms from both the resonant and the non-resonant parts become non-negligible, 
resulting in additional damping rates proportional to $\omega_1^2\tau_c$. Thus, a resonant external drive along $\hat{x}$ on the equilibrium 
magnetization (along $\hat{z}$ at $t=0$), is expected to produce not only a nutation of the magnetization, but also a decay proportional to
$\omega_1^2\tau_c$ of the nutating magnetization.


\section{Discussions on the theoretical approach}

The second order effects of the irradiation appears as shift and damping terms with amplitudes
proportional to $\omega_1^2\tau_c$. 
As such these terms remain in the equation of motion even when
$\omega_{\rms SL} = 0$, an apparently paradoxical result. Although, we have laid down the premise that,
for this derivation, from equation (\ref{semifinal}) and beyond, $\omega_{\rms SL}\not=0$, yet as discussed
below we can still resolve the paradox by carefully checking the other limits whose values we have assumed
to be based on $\omega_{\rms SL}$. At $\omega_{\rms SL} = 0$, the Hilbert space relevant
to the problem would be a direct product of two disjoint Hilbert spaces and complete unitary
dynamics is expected as discussed before. To this end we note that our treatment begins with a choice of $\Delta t$ over
which many instances of the fluctuation have been assumed to take place. After an ensemble average
over the fluctuations and a partial trace over the lattice variables we obtain the final
equation by approximating the coarse-grained derivative over $\Delta t$ by an ordinary time
derivative.  Such an assumption is meaningful only when  $\omega_{\rms SL} \neq 0$ i.e. when the
spin-states and lattice states are part of a common Hilbert space and as such a wide timescale
separation exists in the problem.  Therefore, the choice of setting $\omega_{\rms SL} = 0$ would
naturally be accomplished provided one selects $\Delta t \rightarrow 0$ as well. In fact analogous
treatments often scale $\Delta t$ with $\omega_{\rms SL}$ to unambiguously indicate that $\Delta t$ and
$\omega_{\rms SL}$ are not two independent parameters \cite{Davies74, NamKu16}. It is obvious that instead of 
setting $\Delta t/\tau_c \rightarrow \infty$ if we take the limit $\Delta t \rightarrow 0$, after taking partial
trace over the lattice and dividing both sides of equation (\ref{m2}) by $\Delta t$, we immediately recover
the pure unitary dynamics due to the irradiation, since all second order terms vanish in this limit.

Finally, we note that the Bloch-Siegert shift does not explicitly depend on $\omega_{\rms SL}$, and the
Bloch-Siegert shift has also been experimentally verified \cite{blochsiegert1940, SaWiHaVo10}. 
Thus, if a shift term can exist (verified experimentally), which does not depend 
on the coupling to the bath, then its corresponding decay term (all second order processes usually
appear with a Lamb shift and a corresponding decay) must also exist and would be independent 
of $\omega_{\rms SL}$. 

It is important to delve deeper into the origin of the exponential decay factor,
$\exp(-\vert\tau\vert/\tau_c)$, which appears in all the second order terms of equation (\ref{mf}). The
generic state of a particular spin-bearing molecule, $\vert\psi(t)\rangle$ can be expanded in the
product basis as $\vert\psi(t)\rangle = \sum\limits_{j,k}c_{jk}(t)\vert\chi_j\rangle\otimes\vert\phi_k
\rangle$, 
where $\lbrace\vert\chi_j\rangle\rbrace$ are the eigen-states of $\mathcal{H}_{\rms S}^{\circ}$. Thus
$U_{\rms L}(t_1)$ acting on $\vert\psi(t)\rangle$ introduces random phases into the state-function as
$U_L(t_1)\vert\psi(t)\rangle  =
\sum_{j,k}c_{jk}(t)\exp\left\{-i\int_t^{t_1}dt_2\; f_k(t_2)\right\}
\vert\chi_j\rangle\otimes\vert\phi_k\rangle$
In all the second order terms of equation (\ref{semifinal}), $U_{\rms L}$s appear with time-instances
inherited from the Hamiltonian $H_{\rms eff}$.  In these terms, the external drive acts at time
instants $t_1$ and $t_2$ preserving secularity (net change of quantum numbers to be zero or negligible) 
while the state functions pick-up random phases from the fluctuations through $U_{\rms L}(t_1)U_{\rms
L}^{\dagger}(t_2)$. Therefore, although the drive $H_{\rms S}(t)$ commutes with the fluctuation
$H_{\rms L}(t)$, the random phase thus picked up by the state-functions over the coarse-grained time
interval $\vert t_1 - t_2\vert$ gives rise to a decay after ensemble averaging.

It seems natural that one may move to a frame of $H_{\rms L}(t)$ through a transformation by
$U_{\rms L}$ in which case, only the $H_{\rms SL}$ term would acquire a stochastic nature. But,
$H_{\rms L}(t)$ being a function of time, the important effect of the time-ordering in the
propagation (which essentially leads to the finite second order contribution of the drive) would be
lost in the second order. After all, the usual prescription of the time-dependent perturbation suggests
that only the time-independent part of the Hamiltonian (which is immune to time-ordering) may be
removed by moving to an interaction representation \cite{Dirac58}.

The form
of the Bloch-Siegert shift obtained in the earlier treatments (assuming stroboscopic measurement
protocols) matches with our expression only in the asymptotic limit
\cite{shirley1965,haeberlen1976,madhu2006}. 
Our method enjoys the privilege that the detection does not have to be a stroboscopic measurement.
In any case, having a Hamiltonian which is not 
purely periodic does not strictly permit the application of AHT or Floquet methods. 


\section{Experimental} 
Since $\tau_c \sim 10^{-12}$s in liquids, the application of a transverse drive with $\omega_1 \sim 100$ k-rad/s
in a $500$MHz NMR machine, results in $\omega_{\rms BS}\sim 10^{-5}$rad/s, which is negligible in comparison to the 
on-resonance nutation frequency, $\omega_1$ \cite{samidor85,tanabe78,gilgri72}. Hence, the application of an on-resonance 
($\Delta\omega = 0$) drive to the spin-$\frac{1}{2}$ ensemble under the above conditions, effectively confines the dynamics of 
the magnetization to the $y-z$ plane ($\delta\omega$ is negligible). In this case, all the absorptive parts of the second order 
drive susceptibilities are significant if $\Omega\tau_c < 1$ and the damping coefficients can be approximated as, $\eta_z \approx 2
\omega_1^2\tau_c$, $\eta_x \approx \frac{1}{2}\omega_1^2\tau_c$ and $\eta_y \approx \frac{3}{2}\omega_1^2\tau_c$. 

When $\omega_1> 1/T_1, 1/T_2$  we get a damped nutation in the $y-z$ plane, starting from $\hat{z}$. The nutation frequency is given by 
$\sqrt{\omega_1^2 +\frac{1}{4}(1/T_2- 1/T_1 -\omega_1^2\tau_c/2)^2}$ $\simeq \omega_1$ and the damping rate is 
$(T_1 + T_2) /2T_1T_2 + 7\omega_1^2\tau_c/4$. Since $\tau_c$ is of the order of a few picoseconds in liquids and typically
$\omega_1\sim\,$k-rad/s, the term $\omega_1^2\tau_c$ is usually small in comparison to $(T_1 +
T_2)/2T_1T_2$ \cite{samidor85,tanabe78,gilgri72}. Thus in order to observe its effect, we drive the
system for long enough times (of the order of $100\,$ms).

We have performed all experiments on a Bruker Avance III $11.78$\,T NMR spectrometer at $294$\,K.
The drive strength $\omega_1/2\pi$ has been varied from $3$\,kHz to $20$\,kHz, in steps of $1$\,kHz.
For each drive strength, we have determined the decay rate of nutation, $R_z$ as a function of
nutation period. A plot of $R_z$ as a function of $\omega_1$ is shown in Fig. 2.

The chemical shift of the chosen imine protons is $\sim 8.4$ppm w.r.t. TMS \cite{jberashiha13}. The
choice of the molecule is motivated by its relatively slow isotropic rotation in solution phase due
to its long structure. The $T_1$ and $T_2$ relaxation times for our system, measured using standard
techniques, are $1.34$\,s and $0.81$\,s respectively \cite{furo81,cp54,mg58}.  For each value of
$\nu_1 = \omega_1/2\pi$, we drive the system $n$ times, where $n$ ranges from $1$ to $121$ in steps
of $5$, so that the maximum drive-time does not exceed $500$\,ms. Thus for $\nu_1 = 3$\,kHz the
maximum value of $n$ is $91$. For all other values of $\nu_1$ we use the full range of $n$. This
ensures that imperfections due to finite rise and fall times of the pulses have nearly identical
effects for each $\nu_1$. While $20$\,kHz is approximately close to the maximum power limit of the
spectrometer, we do not go below the $3$\,kHz limit simply because the drive time exceeds $500$\,ms
for a significantly small value of $n$ compared to $121$ thereby leading to distortions. We
calculate $M_z$ after each experimental run from the FID using Plancherel's Theorem \cite{weiner33}.
Taking the natural logarithm of the time series of $M_z$ and fitting the result with a straight line
we obtain $R_z$ for a particular value of $\omega_1$.

The drive strength, $\omega_1$ is not entirely homogeneous throughout the sample volume in practical
cases.  This leads to an additional decay of the measured signal due to the fanning out of the
magnetization components from different parts of the sample, in the $y-z$ plane.  Usually the measured decay of nutation 
is dominated by this first-order dephasing, due to the presence of drive-inhomogeneity in the sample, which obscures
the $\omega_1^2$ dependence of the damping rate. To avoid this we adopt a refocusing scheme such that
$\omega_1 t = 0$ at every instant of measurement $t$. The assumptions made above furnish a solution of the form 
$M_z(t) = a + b\exp[-R_z t]$ and $M_y(t) = 0$, where $R_z = (T_1 + T_2)/2T_1T_2 + \omega_1^2\tau_c$, $a$ and $b$ are 
functions of $M_{\circ}$ (equilibrium magnetization), $T_1$, $T_2$, $\omega_1$ and $\tau_c$ . So we measure the 
decay rates, $R_z$ at various drive strengths, $\omega_1$ to establish the quadratic dependence of the former on the latter. 

\begin{figure}[t] 
\includegraphics[width=5in]{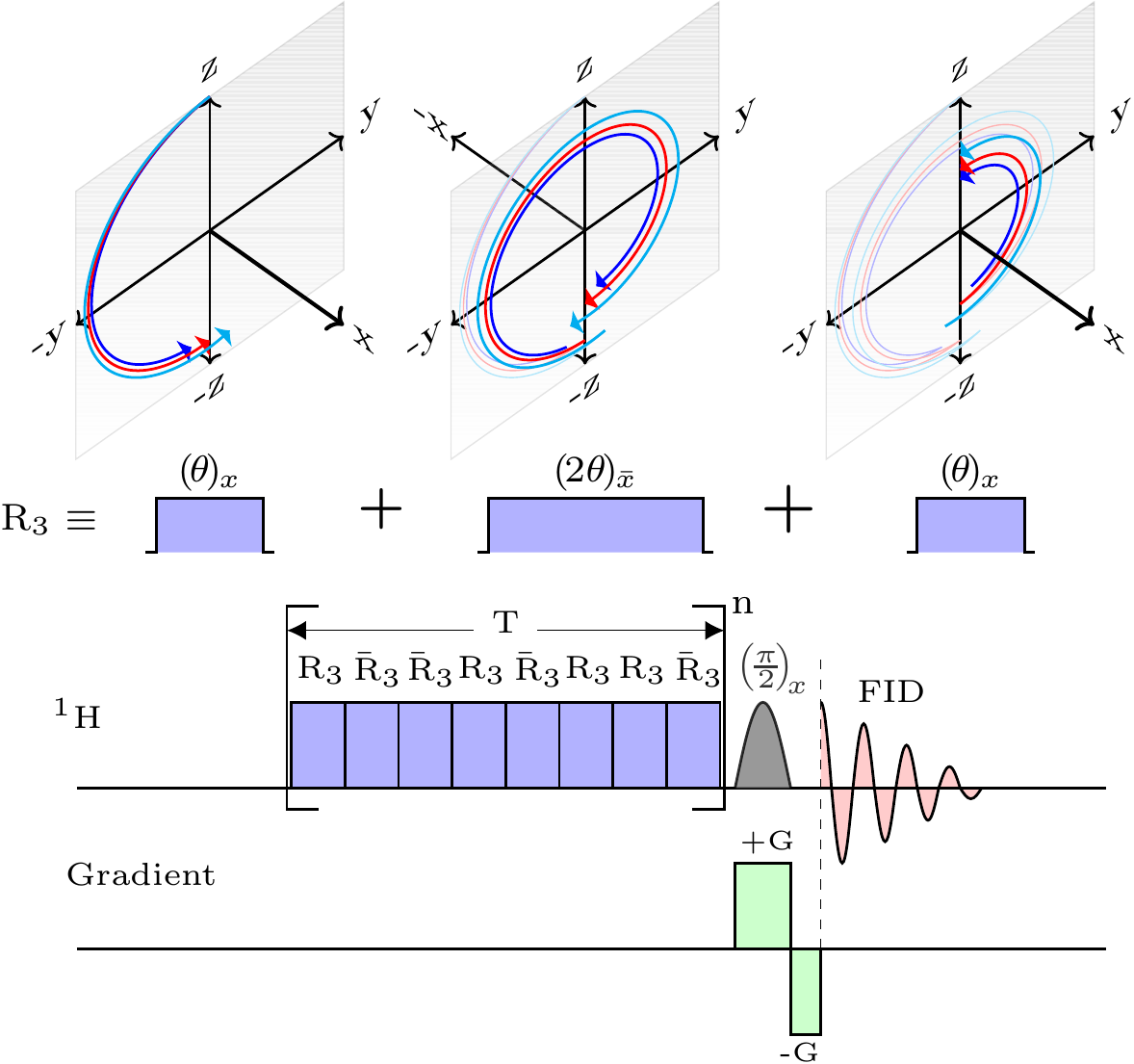}
\caption{\textbf{Refocusing protocol to minimize the effect of drive-inhomogeneity:} Schematic
representation of the refocusing scheme and the pulse sequence used in the experiment.  $\text{R}_3$
is equivalent to an on-resonance $3$-pulse block composed of $\theta, 2\theta, \theta$ pulses along
$\hat{x}, -\hat{x}, \hat{x}$ respectively with $\theta\sim\pi$. $\bar x$ in the figure denotes the
drive applied along $-\hat x$. The 3D diagrams atop the pulses show how the refocusing happens
irrespective of the actual value of the flip angle $\theta$ and hence $\omega_1$. The pulse sequence
below shows the operations performed on the proton ($^1$H) and the gradient channels respectively.
After a recycle delay of $10\, {\rm s} (> 6T_1)$, we employ a super-cycle of R$_3$ inspired by
WALTZ-8 as the drive sequence \cite{shakeefrefre83}. $\bar{\kern1.8pt\rm R}_3$ indicates all pulses
in R$_3$ are applied along inverted axis. $\big[\hspace{0.1cm}\big]^n$ implies that the sequence
within the brackets is repeated $n$ times consecutively and $T$ denotes the length of a super-cycle.
For detection, we have applied a Gaussian selective pulse of flip angle $\sim\pi/2$ in the presence
of the applied gradient $+\text{G}$ to facilitate slice selection using spatial encoding.
Immediately after the slice selection, an opposite compensatory gradient $-\text{G}$ is used to
minimize phase distortion during slice-selection \cite{pavuram16}. The Free Induction Decay (FID)
has been recorded in the absence of any gradient. Magnetization, $M_z$ is measured from the FID as a
function of $n$. From the $M_z$ versus $nT$ data we measure $R_z$ using least-square curve fit.}
\label{refocussing-scheme} 
\end{figure}

An efficient refocusing process should aim to make $M_y(t)$ as small as possible at each measuring
instant $t$.  The simplest possible method to accomplish this involves the application of a
continuous train of pulses with flip-angles $\theta$ and $-\theta$ (i.e. $\omega_1 t = \theta,
-\theta$) alternately with measurements made after even number of pulses.  Nutation by an angle
$-\theta$ is achieved by applying the drive along $-\hat{x}$. Since after an even number of pulses
in the sequence, the net phase of the nutating magnetization components become zero, the resulting
dynamics remains immune to the drive inhomogeneity-induced dephasing.  We choose $\theta\sim\pi$ for
our experiments to minimize precession about $\hat{z}$ (if any, due to incorrect shim profile) which
takes the magnetization out of the $y-z$ plane (during the finite rise and fall times of the pulses).
The $3$-pulse block, $\text{R}_3 = \lbrace\pi, -2\pi, \pi\rbrace$ is more efficient than the simple
$\text{R}_2 = \lbrace\pi, -\pi\rbrace$ block in minimizing the $y$-leakage, $M_y(t)$. This can be
verified by calculating the ratio of the leakage magnetizations, $M_y^{\text{R}_3}(t)/M_y^{2\text{R}_2}(t)$ 
obtained by running the two sequences for the same duration, $t = 4\pi/\omega_1$ (i.e. $\text{R}_2$ is run 
twice consecutively while $\text{R}_3$ is run only once). This ratio turns out to be $-\tanh[\pi\lbrace 1/T_1 + 1/T_2 +
7\omega_1^2\tau_c/2\rbrace/2\omega_1]$. Now, $\omega_1 > 1/T_1,1/T_2$ and $\omega_1^2\tau_c\ll (T_1 +
T_2)/2T_1T_2$ as mentioned before and since for small $\theta$, $\tanh[\theta]\sim\theta$, the ratio
$M_y^{\text{R}_3}(t)/M_y^{2\text{R}_2}(t)$ is negligible. Inspired by the WALTZ-$8$ decoupling
scheme, we find, through simulations with $\tau_c \sim 10^{-12}$s, that the super-cycle $\text{S} =
\text{R}_3\bar{\text{R}}_3\bar{\text{R}}_3\text{R}_3\hspace{0.1cm}\bar{\text{R}}_3\text{R}_3\text{R}_3\bar{\text{R}}_3$, 
(where $\bar{\text{R}}_3 = \lbrace-\pi, 2\pi, -\pi\rbrace$) is more effective than the simple $3$-pulse
block $\text{R}_3$ in minimizing $M_y(t)$ and thus we use it as our driving protocol
\cite{shakeefrefre83}. In order to further eliminate the effect of the diffusion (spins diffusing to
regions having a different $\omega_1$), we select a
thin slice ($\sim 1\,$mm) near the middle of the sample (effective height $\sim 25\,$mm), for
detection, by using a selective Gaussian $\pi/2$ pulse along $\hat{x}$ in presence of an applied
$z$-gradient having strength $\sim 0.05 \text{T}\text{m}^{-1}\ $ \cite{mansfield88, pavuram16}. The duration of
the positive gradient is $2\,$ms whereas that of the compensatory gradient is $1.1\,$ms.  The
experimental protocol is illustrated in Fig. \ref{refocussing-scheme}.
We emphasize that ours is not the traditional spin-locking condition, rather
a refocussed nutation, which we believe is being reported for the first time. We also note that
a small-volume sample chosen near the middle of the tube may remove the requirement for the
slice-selection protocol, but the detection scheme being identical in all experiments, either of
the scheme (small volume or slice-selection) has no bearing on the final outcome which only depends
on the variation of the excitation scheme.

\begin{figure}[t] 
\begin{center} 
\includegraphics[width=5in]{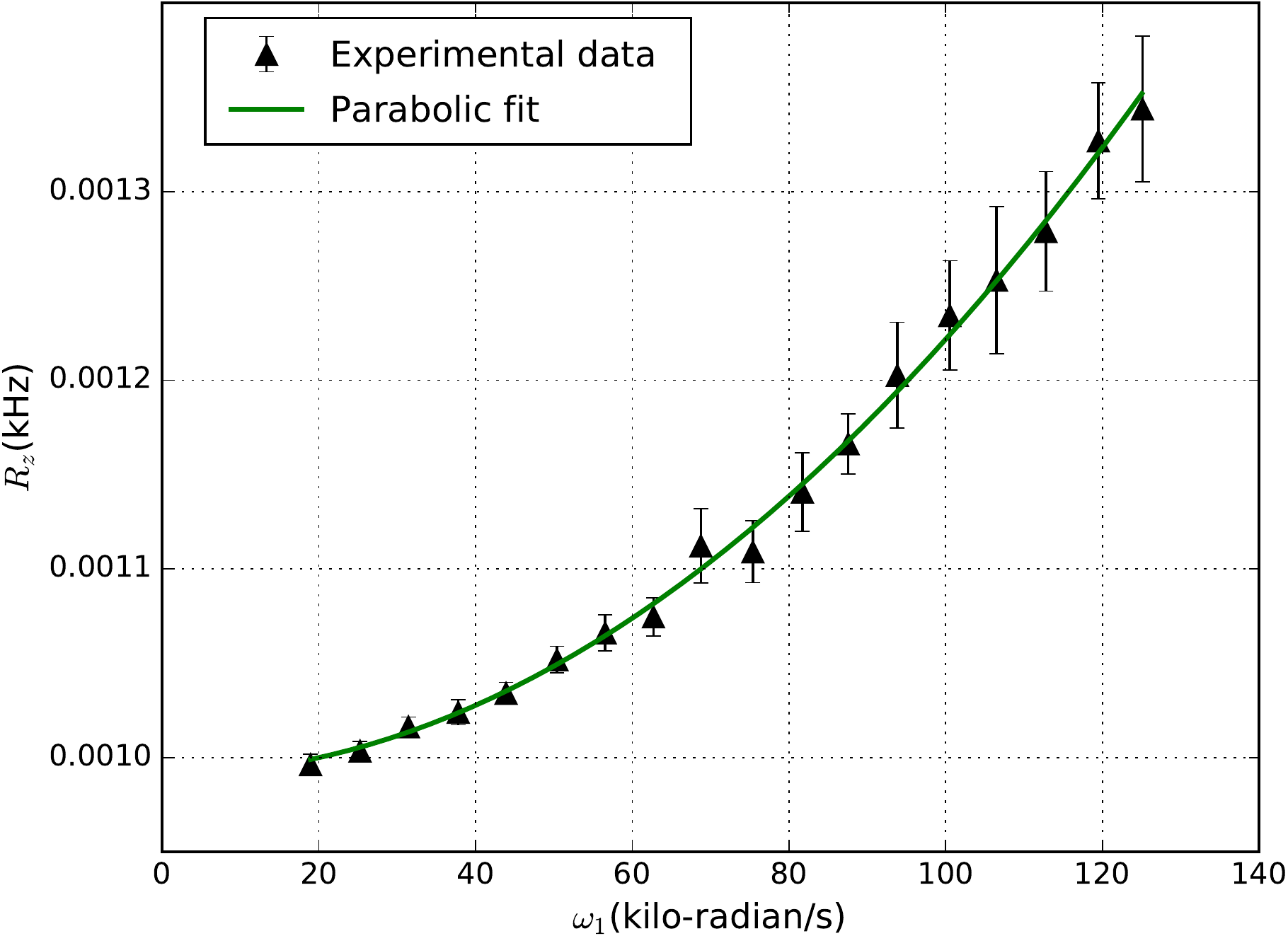}
\caption{\textbf{Experimentally observed damping of nutation due to drive:} Plot of the decay rate
$R_z$ versus the drive strength $\omega_1$.  The upright triangles denote the experimentally
determined values of $R_z$ with the vertical bars denoting the errors in the determination of $R_z$
at 95$\%$ level of confidence. The solid line (color online) shows the parabolic fit of the
experimental data. Since the traditional Bloch Equations, predict $R_z$ to be independent of $\omega_1$, the 
above result unequivocally verifies the existence of the additional decay term originating from the absorptive part 
of the second order drive perturbation.} 
\label{result} 
\end{center} 
\end{figure}

To experimentally detect decay terms proportional to $\omega_1^2$, we choose the singlet imine protons 
in a dilute (millimolar) solution of commercially procured N-(4$'$-methoxybenzylidene)-4-n-butylaniline 
(MBBA) in deuterated dimethyl sulfoxide (DMSO-D$_6$) placed in a static magnetic field along $\hat{z}$, as
our target spin system. The application of an on-resonance ($\Delta\omega\rightarrow 0$) drive along
$\hat{x}$, is shown to result in a damped nutation in the $y-z$ plane (if $\omega_1> 1/T_1, 1/T_2$),
starting from $\hat{z}$, with a damping rate proportional to $\omega_1^2$ as illustrated in Fig.\ref{result}.


\section{Results and Discussions} 

The $R_z$ versus $\omega_1$ data is fit with a parabola of the form $y = a_1 + b_1 x^2$ as shown in
Fig. 2. From the fit we obtain the value of $\tau_c$ to be $1.32\times10^{-11}\,$s. The value of
$a_1$ is found to be $0.99\,$Hz which is same as the value of $(T_1 + T_2)/2T_1T_2$ within
experimental error, the latter being estimated using standard measurement techniques
\cite{furo81,cp54,mg58}. For the $R_z$ versus $\omega_1^2$ behavior, we find fair agreement between
the theoretical prediction and the experimental observation. Hence we infer that our experimental
results confirm the existence of the absorptive counter part of the Bloch-Siegert terms as predicted 
by our master equation (\ref{mf}).

In all the experiments we calibrate our pulses using the fact that a $1.3$db
pulse produces a nutation frequency of $14.25$kHz. The success of the refocusing scheme also relies
on minimal translational diffusion during $\rm{R}_3$ which we presume to be the case since the
duration of $\rm{R}_3$ is of the order of tens of $\mu$s. The maximum drive strength used in our
experiments is $\omega_1/2\pi = 20$kHz (close to the maximum permissible power) whereas the proton Larmor
frequency of the spectrometer is of about $500$MHz. Thus the perturbative expansion used in deriving our
master equation is valid for all practical considerations. 

It is also possible to estimate the second order drive contribution through a spin-locking experiment,
provided the initial magnetization is perfectly aligned along $x$ direction. However, the presence of
the drive inhomogeneity will invariably leave part of the magnetization non-aligned along $x$. These
components will exhibit nutation in the $y-z$ plane and subsequent analysis would be complicated. 
However, the on-resonance refocussed nutation scheme does not suffer from the effects of the drive 
inhomogeneity.

Our equation, as discussed before, allows $\omega_1$ to be a function of time as long as
the timescale of change of $\omega_1$ remains slow compared to that of the fluctuations, which is the case
in all pulsed NMR experiments in liquids. While the pulses are usually separated by sub-$\mu\!\text{s}$
rise-times and the tails, the timescales (of changes in the pulse envelopes) are several orders of
magnitude larger 
than $\tau_c$, thereby the application of this formalism to pulsed NMR is completely within the 
assumed theoretical limits.

If a resonant drive is applied with $\Delta\omega\approx 0$, the 
evolution of the magnetization remains approximately confined to the $y-z$ plane. The departure from the $y-z$ 
plane i.e. the non-fulfillment of the resonant condition, occurs mostly due to incorrect shim conditions 
(static field inhomogeneity). Under a well-shimmed condition, $\Delta\omega\lesssim 1 \text{Hz}$, and
drive strengths $\omega_1\ge 1 \text{kHz}$ -- conditions which we satisfy in our experiments --
the angle of the effective nutation axis is $\tan(\Delta\omega/\omega_1)\sim 10^{-3}\sim 0$
(\textit{w.r.t.} $x$ axis)
for all practical purposes. Therefore, as long as the motion is effectively confined to the $y-z$
plane, change in the pulse phase by a value of $\pi$, would only result in a first order reversal
of the motion of the magnetization while leaving the second order effects unchanged. 
The decay due to departure from the $y-z$ plain is expected to scale
with the offset frequency $\Delta\omega$ and not with $\omega_1^2$ and as such cannot be used
to explain our experimental findings.


\section{Conclusion} Our results reveal that the finite memory of the bath plays a central role in
giving rise to time-nonlocal second-order complex susceptibility terms, having both dispersive and
absorptive parts, from the external drive applied to the system. While the imaginary part of the
susceptibility manifests itself as a dispersive shift term in the dynamics, its corresponding
absorptive part provides a damping in addition to usual relaxation terms.  The shift term due to
resonant part of the drive is negligible, but the same due to the non-resonant part of the drive
appears as the Bloch-Siegert shift under an asymptotic limit. The correction terms from the drive
are extremely small yet measurable under suitable conditions, in the context of the solution-state
NMR spectroscopy. For solid-state spectroscopy, such terms may not be negligible and provides a
natural explanation for saturation processes. At extreme motional narrowing limit ($\tau_c
\rightarrow 0$), the equation (\ref{mf}) predicts the expected unitary evolution of the system due
to the drive, in the form of pure nutation.  Our choice of $\delta$-correlated fluctuations lead to
an exponential memory function; although other noise models may also serve the purpose.  Since
$\omega_1^2\tau_c \ll 1$, this damping decreases for smaller $\tau_c$ and hence it is envisaged that
a bath with stronger fluctuations (towards the extreme motional narrowing limit) would make the
system more immune to drive-induced damping and the shifts, which may have important consequences
for ensemble quantum computing.

\section{Acknowledgments}
We acknowledge IISER Kolkata for providing the necessary funding and for providing access to the
central NMR facility. A.C. would like to thank Anirban Mukherjee for insightful discussions and
Centre for Scientific and Industrial Research (CSIR) India, for a senior research fellowship.

\appendix
\section{Construction of the finite propagator}

Following the description of the system and the lattice laid out in the main manuscript and the notations
introduced therein, we intend to construct a finite propagator $U(t_1)$, valid for the coarse-grained time interval, $t_1 \in 
[t, t + \Delta t]$. The coarse-graining time $\Delta t$ is such that the contribution of $H_{\rms eff}$ in $U(t_1)$ can be linearized
and only the leading first order terms are retained. On the contrary, many instances of the fluctuation take place within $\Delta t$
and as such we retain all possible higher order terms of $H_{\rms L}$. The explicit construction of the finite-time propagator begins 
from the Schr{\"o}dinger equation:

\be\label{SSch}
    \frac{d}{dt}U(t) = -i\,H(t)\,U(t)
\ee
and its formal solution in the domain $[t, t_1]$ (with $t_1> t$):

\bea\label{SNeu}
       U(t_1) & = & \mathbbm{1} - i\,\int\limits_t^{t_1}dt_2\,H(t_2)\,U(t_2)\nonumber\\
              & = & \mathbbm{1} - i\,\int\limits_t^{t_1}dt_2\,H_{\rms eff}(t_2)\,U(t_2) - i\,\int\limits_t^{t_1}dt_2\,H_{\rms L}(t_2)\,U(t_2).
\eea
Since $t_1 \in [t, t + \Delta t]$ by assumption, the interval $(t_1 - t) \ll 1/\omega_1,1/\omega_{\rms SL}$ and as such further
propagation due to $H_{\rms eff}$ can be neglected on the r.h.s of equation (\ref{SNeu}). Thus collecting all the remaining terms
in the r.h.s. of equation (\ref{SNeu}) we get a finite propagator with a leading linear order term in $H_{\rms eff}$ of the form

\be\label{SProp1}
       U(t_1) \approx \mathbbm{1} - i\,\int\limits_t^{t_1}dt_2\,H_{\rms eff}(t_2)\,U_{\rms L}(t_2) - i\,\int\limits_t^{t_1}dt_2\,H_{\rms L}(t_2)
                                     \,U_{\rms L}(t_2).
\ee
Combining the last term in the \textit{r.h.s.} with the identity
we can rewrite the above equation as,
\be\label{SProp2}
       U(t_1) \approx U_{\rms L}(t_1) - i\,\int\limits_t^{t_1}dt_2\,H_{\rms eff}(t_2)\,U_{\rms L}(t_2).
\ee


\section{Emergence of the memory of the bath from the fluctuations}

Following the usual practice, we too assume that at the beginning of the coarse-graining interval the full density matrix has the factorized form

\be
     \rho(t) = \rho_{\rms S}(t)\otimes\rho_{\rms L}^{\rms eq},
\ee
where $\rho(t) = \overline{\widetilde \rho(t)}$ and $\rho_{\rms L}^{\rms eq} = \text{exp}(-\beta\mathcal{H}_{\rms L}^{\circ})/
\mathcal{Z}_{\rms L}$ denotes the equilibrium density matrix of the lattice \cite{cotandurogryn04,wangblo53,brepet02}. 

We thus have, 
\begin{equation}\label{Sddm1}
    \overline{U_{\rms L}(t_1)\widetilde\rho(t) U_{\rms L}^{\dagger}(t_2)}  =  \rho_S(t)\otimes\sum_j\frac{e^{-\beta\omega_j}}
                                                                              {\mathcal{Z}_L}\vert\phi_j\rangle\langle\phi_j\vert\overline
                                                                              {\exp \big\lbrace -i\int_{t}^{t_1} dt_3\,f_j(t_3) 
                                                                              + i\int_{t}^{t_2} dt_4\,f_j(t_4)\big\rbrace}. 
\end{equation}

In the above expression $\omega_j$ denotes the eigen-value of $\mathcal{H}_{\rms L}^{\circ}$ corresponding to
$\vert\phi_j\rangle$ and we have made use of the fact that $H_{\rms L}(t)$ and thus the propagators $U_{\rms L}(t_1)$ 
are independent of the initial distribution of the lattice states $\forall t_1\geq t$.

Thus we obtain, 

\be\label{Sddm2}
           \overline{U_{\rms L}(t_1)\widetilde\rho(t) U_{\rms L}^{\dagger}(t_2)}  = \rho_{\rms S}(t)\otimes\rho_{\rms L}^{\rms eq}\exp\big
                                                                                    (-\frac{1}{2}\kappa^2\vert t_1 - t_2\vert\big).\\
\ee 

In deriving the above, we have used the cumulant expansion for Gaussian stochastic processes with usual $\delta$-correlation in time,
for which only the terms upto the second cumulant survive. A further assumption of zero mean (as in
our model) leaves only the exponentially decaying factor, $\exp\big(-\frac{1}{2}\kappa^2\vert t_1 - t_2\vert\big)$.


\section{Explicit derivation of the drive-contributions in the Bloch Equations}

The measured magnetization components for the spin-$1/2$ ensemble are given by $M_{\alpha}(t) = \text{Tr}_{\rms S}[
F_{\rms \alpha}^{\rms R}(t)\rho_{\rms S}(t)]$, $\alpha\in \lbrace x,y,z\rbrace$. Thus, using the master equation derived in the main 
manuscript and the observables defined using equations (11) in the manuscript, we obtain,

\bea\label{Sde1}
   \frac{d}{dt}M_{\rms \alpha}(t) & = & \text{Tr}_{\rms S}\Big[\Big\lbrace\frac{d}{dt}F_{\rms \alpha}^{\rms R}(t)\Big\rbrace\rho_{\rms S}(t)\Big] +
                                              \text{Tr}_{\rms S}\Big[F_{\rms \alpha}^{\rms R}(t)\Big\lbrace\frac{d}{dt}\rho_{\rms S}(t)\Big\rbrace\Big]\nonumber\\
                                  & = & \text{Tr}_{\rms S}\Big[\Big\lbrace\frac{d}{dt}F_{\rms \alpha}^{\rms R}(t)\Big\rbrace\rho_{\rms S}(t)\Big] 
                                        -i\,\text{Tr}_{\rms S}\Big\lbrace {\rm Tr}_{\rms L}\!\left[H_{\rms eff}(t),\, \rho_{\rms S}(t)\otimes\rho_{\rms L}
                                        ^{\rms eq}\right]^{\rms sec}F_{\rms \alpha}^{\rms R}(t)\Big\rbrace \nonumber\\
                                  &   & - \int\limits_{0}^{\infty}\kern-2pt d\tau\;\text{Tr}_{\rms S}\Big\lbrace{\rm Tr}_{\rms L}\!\left[H_{\rms eff}(t),
                                          \,\left[H_{\rms eff}(t-\tau), \rho_{\rms S}(t)\otimes\rho_{\rms L}^{\rms eq}\right]\right]^{\rms sec}
                                          e^{-{\vert\tau\vert}/{\tau_c}}F_{\rms \alpha}^{\rms R}(t)\Big\rbrace.
\eea
The first term in the r.h.s.\,of the last equation above is easy to evaluate. For the dynamical equation for $M_{\rms y}(t)$, this term becomes

\bea\label{Sde11y}
      \text{Tr}_{\rms S}\Big[\Big\lbrace\frac{d}{dt}F_{\rms y}^{\rms R}(t)\Big\rbrace\rho_{\rms S}(t)\Big] & = &\text{Tr}_{\rms S}\Big[
                                              \frac{1}{2i}\frac{d}{dt}\big\lbrace I_+e^{-i\Delta\omega t} - I_-e^{i\Delta\omega t}\big\rbrace\rho_{\rms S}(t)
                                              \Big]\nonumber\\
                                                                                                             & = & -\Delta\omega\text{Tr}_{\rms S}\Big[
                                              \frac{1}{2}\big\lbrace I_+e^{-i\Delta\omega t} + I_-e^{i\Delta\omega t}\big\rbrace\rho_{\rms S}(t)\Big]\nonumber\\
                                                                                                             & = & -\Delta\omega\,M_{\rms x}(t).
\eea
Similarly, for the other magnetization components we have from the first term in the r.h.s of equation (\ref{Sde1}),

\be\label{Sde11z}
      \text{Tr}_{\rms S}\Big[\Big\lbrace\frac{d}{dt}F_{\rms z}^{\rms R}(t)\Big\rbrace\rho_{\rms S}(t)\Big]  = 0                                          
\ee
and

\be\label{Sde11x}
      \text{Tr}_{\rms S}\Big[\Big\lbrace\frac{d}{dt}F_{\rms x}^{\rms R}(t)\Big\rbrace\rho_{\rms S}(t)\Big]  = \Delta\omega\,M_{\rms y}(t).                                         
\ee

\subsection{First order drive contribution}

The condition $\text{Tr}_{\rms L}[H_{\rms SL}(t),\rho_{\rms S}(t)\otimes\rho_{\rms L}^{\rms eq}] = 0$ ensures that only the external drive, 
$H_{\rms S}(t)$ contributes in the second term on the r.h.s.\,of equation (\ref{Sde1}). Of this only the secular terms (co-rotating terms 
i.e. terms with frequency $\Delta\omega$ in the interaction representation) are non-negligible. Thus we have

\bea\label{Sde12}
   -i\,\text{Tr}_{\rms S}\Big\lbrace {\rm Tr}_{\rms L}\!\left[H_{\rms S}(t),\, \rho_{\rms S}(t)\otimes\rho_{\rms L}
                                        ^{\rms eq}\right]^{\rms sec}F_{\rms \alpha}^{\rms R}(t)\Big\rbrace & = &  -i\,\text{Tr}_{\rms S}\Big\lbrace 
                                         \omega_1{\rm Tr}_{\rms L}\!\left[F_{\rms x}^{\rms C}(t) + F_{\rms x}^{\rms R}(t),\, \rho_{\rms S}(t)
                                         \otimes\rho_{\rms L}^{\rms eq}\right]^{\rms sec}F_{\rms \alpha}^{\rms R}(t)\Big\rbrace\nonumber\\
                                                                                                             & = &  -i\,\text{Tr}_{\rms S}\Big\lbrace 
                                         \omega_1\!\left[F_{\rms x}^{\rms C}(t) + F_{\rms x}^{\rms R}(t),\, \rho_{\rms S}(t)\right]^{\rms sec}F_{\rms 
                                         \alpha}^{\rms R}(t)\Big\rbrace\nonumber\\
                                                                                                             & = & -i\,\text{Tr}_{\rms S}\Big\lbrace 
                                         \omega_1\!\left[F_{\rms x}^{\rms R}(t),\, \rho_{\rms S}(t)\right] F_{\rms \alpha}^{\rms R}(t)\Big\rbrace\nonumber\\
                                                                                                             & = & -i\,\text{Tr}_{\rms S}\Big\lbrace 
                                         \omega_1\!\left[F_{\rms \alpha}^{\rms R}(t),\, F_{\rms x}^{\rms R}(t)\right]\rho_{\rms S}(t)\Big\rbrace,
\eea
where in the second-last step we have removed the non-secular counter-rotating term, $F_{\rms x}^{\rms C}(t)$ and hence the superscript `sec' is
dropped thereafter. In the above derivation we have used the fact that ${\rm Tr}_{\rms L}[\rho_L^{\rm eq}] = 1$. For evaluating the commutators of 
$F_{\rms \alpha}^{\rms R}(t)$ we note that


\bea\label{Comm}
    \Big[F_{\rms \alpha_k}^{\rms R}(t) , F_{\rms \alpha_m}^{\rms R}(t)\Big] & = & \Big[e^{-i\Delta\omega t I_z}I_{\rms \alpha_k}e^{i\Delta\omega t I_z},
                                                                                        e^{-i\Delta\omega t I_z}I_{\rms \alpha_m}e^{i\Delta\omega t I_z}\Big]\nonumber\\
                                                                                & = & e^{-i\Delta\omega t I_z}\Big[I_{\rms \alpha_k},I_{\rms \alpha_m}\Big]
                                                                                      e^{i\Delta\omega t I_z}\nonumber\\
                                                                                & = & i\,\varepsilon_{pkm}\,e^{-i\Delta\omega t I_z}I_{\rms \alpha_p}e^{i\Delta\omega t I_z}
                                                                                      \nonumber\\
                                                                                & = & i\,\varepsilon_{pkm}\,F_{\rms \alpha_p}^{\rms R}(t),
\eea
where, $\lbrace k,m,p \rbrace \in \lbrace 1,2,3 \rbrace$, $\alpha_1 = x$, $\alpha_2 = y$ and $\alpha_3 = z$ and $\varepsilon_{pkm}$ is the Levi-Civita symbol.
Substituting the expression for $F_{\rms \alpha}^{\rms R}$ in the last line of the above expression one can readily find the 
first-order drive contribution in the dynamics of $M_{\rms \alpha}(t)$. For example, in the dynamics of $M_{\rms y}(t)$, we have

\bea\label{Sde12y}
     -i\,\text{Tr}_{\rms S}\Big\lbrace {\rm Tr}_{\rms L}\!\left[H_{\rms S}(t),\, \rho_{\rms S}(t)\otimes\rho_{\rms L}
                                        ^{\rms eq}\right]^{\rms sec}F_{\rms y}^{\rms R}(t)\Big\rbrace & = & -i\,\text{Tr}_{\rms S}\Big\lbrace 
                                         \omega_1\!\left[F_{\rms y}^{\rms R}(t),\, F_{\rms x}^{\rms R}(t)\right]\rho_{\rms S}(t)\Big\rbrace\nonumber\\
                                                                                                        & = & -i\,\text{Tr}_{\rms S}\Big\lbrace 
                                         \omega_1\!(-i)F_{\rms z}^{\rms R}\rho_{\rms S}(t)\Big\rbrace\nonumber\\
                                                                                                        & = & -\omega_1\,M_{\rms z}(t).
\eea

\subsection{Second order drive contribution}
In the third expression on the r.h.s.\,of equation (\ref{Sde1}), the cross-terms between $H_{\rms SL}(t)$ and $H_{\rms S}(t)$ vanish due to the
condition $\text{Tr}_{\rms L}[H_{\rms SL}(t),\rho_{\rms S}(t)\otimes\rho_{\rms L}^{\rms eq}] = 0$. Only contributions in this expression comes
from the self-terms of $H_{\rms SL}(t)$ and $H_{\rms S}(t)$. We assume a generic form of the coupling Hamiltonian for the spin-$\frac{1}{2}$ ensemble,
following Wangsness and Bloch, which leads to the relaxation terms proportional to $1/T_1$ and $1/T_2$ with the equilibrium magnetization $M_{\circ}$ as
shown in their work (Lamb-shift terms being neglected) \cite{wangblo53}. Since the calculations involved follow exactly that of Bloch and Wangsness
and is not the main focus of this work, we shall assume the existence of these relaxation terms in our equations of motion without elaborating further. The 
second-order drive contribution is given by

\begin{align}\label{Sde13}
    & -\int\limits_{0}^{\infty}\kern-2pt d\tau\;\text{Tr}_{\rms S}\Big\lbrace{\rm Tr}_{\rms L}\!\big[H_{\rms S}(t),
                                          \,\big[H_{\rms S}(t-\tau), \rho_{\rms S}(t)\otimes\rho_{\rms L}^{\rms eq}\big]\big]^{\rms sec}
                                          e^{-{\vert\tau\vert}/{\tau_c}}F_{\rms \alpha}^{\rms R}(t)\Big\rbrace \nonumber\\
     \nonumber\\
    & = -\omega_1^2\int\limits_{0}^{\infty}\kern-2pt d\tau\;\text{Tr}_{\rms S}\Big\lbrace\big[F_{\rms x}^{\rms C}(t) + F_{\rms x}^{\rms R}(t),
                                          \,\big[F_{\rms x}^{\rms C}(t-\tau) + F_{\rms x}^{\rms R}(t-\tau), \rho_{\rms S}(t)\big]\big]^{\rms sec}
                                            e^{-{\vert\tau\vert}/{\tau_c}}F_{\rms \alpha}^{\rms R}(t)\Big\rbrace\nonumber\\
    & = -\frac{1}{4}\omega_1^2\int\limits_{0}^{\infty}\kern-2pt d\tau\;\text{Tr}_{\rms S}\Big\lbrace \big[\big(I_+e^{i\Omega t} + I_-e^{-i\Omega t}\big) 
                                            + \big(I_+e^{-i\Delta\omega t} + I_-e^{i\Delta\omega t}\big), \nonumber\\
    & \hspace{2cm}\big[\big(I_+e^{i\Omega (t-\tau)} + I_-e^{-i\Omega (t-\tau)}\big) + \big(I_+e^{-i\Delta\omega (t-\tau)} + I_-e^{i\Delta\omega 
                                          (t-\tau)}\big), \rho_{\rms S}(t) \big]\big]^{\rms sec}F_{\rms \alpha}^{\rms R}(t)\Big\rbrace 
                                          e^{-{\vert\tau\vert}/{\tau_c}},
\end{align}
where we have again used the fact that ${\rm Tr}_{\rms L}[\rho_L^{\rm eq}] = 1$.

To determine secular terms in the above expression we follow the analysis prescribed by Cohen-Tannoudji \textit{et. al}
\cite{cotandurogryn04}. We note that the terms oscillating with $\Delta\omega$ are not averaged out during the interval
$\Delta t$.
Thus the secular approximation retains the full self-term of $F_{\rms x}^{\rms R}(t)$
in the second-order of the drive. On the other hand all cross-terms between $F_{\rms x}^{\rms C}(t)$ and $F_{\rms x}^{\rms R}(t)$ become negligible
in the secular limit. In the self-terms of $F_{\rms x}^{\rms C}(t)$, only the cross-commutators between $I_+$ and $I_-$ survive the secular integration. Thus
we finally arrive at the following form of the secular second-order drive contributions:

\begin{align}\label{Sde131}
      & - \int\limits_{0}^{\infty}\kern-2pt d\tau\;\text{Tr}_{\rms S}\Big\lbrace{\rm Tr}_{\rms L}\!\big[H_{\rms S}(t),
                                          \,\big[H_{\rms S}(t-\tau), \rho_{\rms S}(t)\otimes\rho_{\rms L}^{\rms eq}\big]\big]^{\rms sec}
                                          e^{-{\vert\tau\vert}/{\tau_c}}F_{\rms \alpha}^{\rms R}(t)\Big\rbrace \nonumber\\
      & = -\frac{1}{4}\omega_1^2\int\limits_{0}^{\infty}\kern-2pt d\tau\;\text{Tr}_{\rms S}\Big\lbrace \big[ I_+e^{i\Omega t},\big[I_-e^{-i\Omega (t-\tau)},
                                           \rho_{\rms S}(t) \big]\big]F_{\rms \alpha}^{\rms R}(t) + \big[ I_-e^{-i\Omega t},\big[I_+e^{i\Omega (t-\tau)},
                                           \rho_{\rms S}(t) \big]\big]F_{\rms \alpha}^{\rms R}(t)\nonumber\\
      &  \hspace{2cm} + \big[\big(I_+e^{-i\Delta\omega t} + I_-e^{i\Delta\omega t}\big),\big[\big(I_+e^{-i\Delta\omega (t-\tau)} + I_-e^{i\Delta\omega (t-\tau)}\big),
                                           \rho_{\rms S}(t) \big]\big]F_{\rms \alpha}^{\rms R}(t)\Big\rbrace e^{-{\vert\tau\vert}/{\tau_c}}\nonumber\\
      & = 
-\frac{1}{4}\omega_1^2 \Big[\int\limits_{0}^{\infty}\kern-2pt d\tau\,e^{i\Omega\tau}e^{-{\vert\tau\vert}/{\tau_c}}\Big]\;\text{Tr}_{\rms S}
 \Big\lbrace\big[I_-,\big[I_+,F_{\rms \alpha}^{\rms R}(t)\big]\big]\rho_{\rms S}(t)\Big\rbrace
-\frac{1}{4}\omega_1^2\Big[\int\limits_{0}^{\infty}\kern-2pt d\tau\,e^{-i\Omega\tau}e^{-{\vert\tau\vert}/{\tau_c}}\Big]\;\text{Tr}_{\rms S}
 \Big\lbrace\big[I_+,\big[I_-,F_{\rms \alpha}^{\rms R}(t)\big]\big]\rho_{\rms S}(t)\Big\rbrace\nonumber\\
      & \hspace{0.5cm} - \omega_1^2\int\limits_{0}^{\infty}\kern-2pt d\tau\;\text{Tr}_{\rms S}\Big\lbrace \big[F_{\rms x}^{\rms R}(t-\tau),\big[F_{\rms x}^{\rms R}(t),
                                           F_{\rms \alpha}^{\rms R}(t)\big]\big]\rho_{\rms S}(t)\Big\rbrace e^{-{\vert\tau\vert}/{\tau_c}}\nonumber\\
      & = 
-\frac{1}{4}\omega_1^2\Gamma(\Omega)\text{Tr}_{\rms S}\Big\lbrace\big[I_-,\big[I_+,F_{\rms \alpha}^{\rms R}(t)\big]\big]\rho_{\rms S}(t)\Big\rbrace
-\frac{1}{4}\omega_1^2\Gamma^*(\Omega)\text{Tr}_{\rms S}\Big\lbrace\big[I_+,\big[I_-,F_{\rms \alpha}^{\rms R}(t)\big]\big]\rho_{\rms S}(t)\Big\rbrace\nonumber\\
      & \hspace{0.5cm} - \omega_1^2\int\limits_{0}^{\infty}\kern-2pt d\tau\;\text{Tr}_{\rms S}\Big\lbrace \big[F_{\rms x}^{\rms R}(t-\tau),\big[F_{\rms x}^{\rms R}(t),
                                           F_{\rms \alpha}^{\rms R}(t)\big]\big]\rho_{\rms S}(t)\Big\rbrace e^{-{\vert\tau\vert}/{\tau_c}},
\end{align}
where $\Gamma (\Omega) = \Big[\int\limits_{0}^{\infty}\kern-2pt d\tau\,e^{i\Omega\tau}e^{-{\vert\tau\vert}/{\tau_c}}\Big]$ and hence its complex conjugate,
$\Gamma^* (\Omega) = \Big[\int\limits_{0}^{\infty}\kern-2pt d\tau\,e^{-i\Omega\tau}e^{-{\vert\tau\vert}/{\tau_c}}\Big]$. $\Gamma (\Omega)$ is a complex Lorentzian 
centered at $\Omega$ having real and imaginary parts given by

\be\label{BS K-K}
     \Gamma (\Omega) = \frac{\tau_c}{1 + \Omega^2\tau_c^2} + i\frac{\Omega\tau_c^2}{1 + \Omega^2\tau_c^2},
\ee
which are Kramers-Kronig pairs. Substituting the expression for $F_{\rms \alpha}^{\rms R}$ in the last line of the equation (\ref{Sde131}) we find the full 
second-order drive contribution in the dynamics of $M_{\rms \alpha}(t)$. Again as an example, the second-order contribution of the drive in the dynamical equation of $M_{\rms y}(t)$ is obtained after substituting $F_{\rms y}^{\rms R}$ in place of $F_{\rms \alpha}^{\rms R}$ and is given by:

\begin{align}\label{Sde13y}
     &  -\frac{1}{4}\omega_1^2\Gamma^*(\Omega)\text{Tr}_{\rms S}\Big\lbrace\frac{1}{2i}\big[I_+,\big[I_-,I_+\big]\big]\,e^{-i\Delta\omega t}\rho_{\rms S}(t)\Big\rbrace
         -\frac{1}{4}\omega_1^2\Gamma(\Omega)\text{Tr}_{\rms S}\Big\lbrace -\frac{1}{2i}\big[I_-,\big[I_+,I_-\big]\big]\,e^{i\Delta\omega t}\rho_{\rms S}(t)\Big\rbrace
         \nonumber\\
     & \hspace{0.5cm} - \omega_1^2\int\limits_{0}^{\infty}\kern-2pt d\tau\;\text{Tr}_{\rms S}\Big\lbrace i\big[\big(I_+e^{-i\Delta\omega (t-\tau)} + I_-e^{i\Delta\omega 
                                          (t-\tau)}\big),I_z\big]\rho_{\rms S}(t)\Big\rbrace e^{-{\vert\tau\vert}/{\tau_c}}\nonumber\\
     & = -\frac{1}{4i}\omega_1^2\Big[\frac{\tau_c}{1 + \Omega^2\tau_c^2} - i\frac{\Omega\tau_c^2}{1 + \Omega^2\tau_c^2}\Big]\text{Tr}_{\rms S}[I_+e^{-i\Delta\omega t}
                                          \rho_{\rms S}(t)] +\frac{1}{4i}\omega_1^2\Big[\frac{\tau_c}{1 + \Omega^2\tau_c^2} + i\frac{\Omega\tau_c^2}{1 + \Omega^2\tau_c^2}
                                          \Big]\text{Tr}_{\rms S}[I_-e^{i\Delta\omega t}\rho_{\rms S}(t)\big]\nonumber\\
     & \hspace{0.5cm} -\frac{1}{2i}\omega_1^2\Big[\frac{\tau_c}{1 + \Delta\omega^2\tau_c^2} - i\frac{\Delta\omega\tau_c^2}{1 + \Delta\omega^2\tau_c^2}\Big]\text{Tr}_{\rms S}
                                          [I_+e^{-i\Delta\omega t}\rho_{\rms S}(t)] +\frac{1}{2i}\omega_1^2\Big[\frac{\tau_c}{1 + \Delta\omega^2\tau_c^2} + i
                                          \frac{\Delta\omega\tau_c^2}{1 + \Delta\omega^2\tau_c^2}\Big]\text{Tr}_{\rms S}[I_-e^{i\Delta\omega t}\rho_{\rms S}(t)
                                          \big].
\end{align}
Simplifying the above expression we have,

\begin{align}\label{Sde13y1}
     & -\frac{1}{2}\omega_1^2\Big[\frac{\tau_c}{1 + \Omega^2\tau_c^2}\Big]\text{Tr}_{\rms S}\big[\frac{1}{2i}\big(I_+e^{-i\Delta\omega t} - I_-e^{i\Delta\omega t}\big)
                                         \rho_{\rms S}(t)\big] -\omega_1^2\Big[\frac{\tau_c}{1 + \Delta\omega^2\tau_c^2}\Big]\text{Tr}_{\rms S}\big[\frac{1}{2i}
                                         \big(I_+e^{-i\Delta\omega t} - I_-e^{i\Delta\omega t}\big)\rho_{\rms S}(t)\big]\nonumber\\
     \nonumber\\
     & \hspace{0.2cm} + \frac{1}{2}\omega_1^2\Big[\frac{\Omega^2\tau_c^2}{1 + \Omega^2\tau_c^2}\Big]\text{Tr}_{\rms S}\big[\frac{1}{2}\big(I_+e^{-i\Delta\omega t} + 
                                         I_-e^{i\Delta\omega t}\big)\rho_{\rms S}(t)\big] + \omega_1^2\Big[\frac{\Delta\omega^2\tau_c^2}{1 + \Delta\omega^2\tau_c^2}\Big]
                                         \text{Tr}_{\rms S}\big[\frac{1}{2}\big(I_+e^{-i\Delta\omega t} + I_-e^{i\Delta\omega t}\big)\rho_{\rms S}(t)\big]\nonumber\\
     \nonumber\\
     & = -\frac{1}{2}\omega_1^2\Big[\frac{\tau_c}{1 + \Omega^2\tau_c^2}\Big]\text{Tr}_{\rms S}\big[F_{\rms y}^{\rms R}(t)\rho_{\rms S}(t)\big] 
                   -\omega_1^2\Big[\frac{\tau_c}{1 + \Delta\omega^2\tau_c^2}\Big]\text{Tr}_{\rms S}\big[F_{\rms y}^{\rms R}(t)\rho_{\rms S}(t)\big]\nonumber\\
     \nonumber\\
     & \hspace{0.2cm} + \frac{1}{2}\omega_1^2\Big[\frac{\Omega^2\tau_c^2}{1 + \Omega^2\tau_c^2}\Big]\text{Tr}_{\rms S}\big[F_{\rms x}^{\rms R}(t)\rho_{\rms S}(t)\big] 
                      + \omega_1^2\Big[\frac{\Delta\omega^2\tau_c^2}{1 + \Delta\omega^2\tau_c^2}\Big]\text{Tr}_{\rms S}\big[F_{\rms x}^{\rms R}(t)\rho_{\rms S}(t)
                      \big]\nonumber\\
     \nonumber\\
     & = -\eta_y M_{\rms y}(t) + (\omega_{\rms BS} + \delta\omega) M_{\rms x}(t),
\end{align}
where,

\be\label{SBS1}
   \omega_{\rms BS} = \frac{1}{2}\Big(\frac{\omega_1^2\Omega\tau_c^2}{1 + \Omega^2\tau_c^2}\Big),
\ee

\be\label{Sshift}
   \delta\omega = \frac{\omega_1^2\Delta\omega\tau_c^2}{1 + \Delta\omega^2\tau_c^2}
\ee
and

\be\label{Sdamp}
   \eta_y  =  \omega_1^2\Big[\frac{1}{2}\Big(\frac{\tau_c}{1 + \Omega^2\tau_c^2}\Big) + \frac{\tau_c}{1 + \Delta\omega^2\tau_c^2}\Big].
\ee

The second-order drive contributions in the dynamics of the other magnetization components follow in a similar way.

\end{document}